\documentclass[reprint,superscriptaddress,nofootinbib,amsmath,amssymb,aps,prd]{revtex4-2}

\usepackage[utf8]{inputenc}
\usepackage[T1]{fontenc}
\usepackage[colorlinks=true, allcolors=Blue]{hyperref}

\usepackage{microtype}

\usepackage{mathtools, booktabs}

\usepackage{graphicx}

\usepackage[dvipsnames]{xcolor}
\usepackage[capitalise]{cleveref}
\usepackage{tensor}
\usepackage{enumerate}

\usepackage{mathrsfs, bm, tensor}

\def\K{\mathcal{K}}
\def\cL{\mathcal{L}}

\newcommand{\hnabla}{\overset{\mathrm{h}}{\nabla}{}}
\newcommand{\vnabla}{\overset{\mathrm{v}}{\nabla}{}}

\numberwithin{equation}{section}

\begin{document}

\title{Conserved quantities and integrability for massless spinning particles in general relativity}

\author{Lars Andersson}
\email{lars.andersson@uni-potsdam.de}
\affiliation{University of Potsdam, Department of Mathematics, Karl-Liebknecht-Str. 24-25, 14476 Potsdam, Germany}

\author{Finnian Gray}
\email{finnian.gray@univie.ac.at}
\affiliation{University of Vienna, Faculty of Physics, Boltzmanngasse~5, 1090 Vienna, Austria}

\author{Marius A. Oancea}
\email{marius.oancea@univie.ac.at}
\affiliation{University of Vienna, Faculty of Physics, Boltzmanngasse~5, 1090 Vienna, Austria}

\begin{abstract}
In general relativity, the dynamics of spinning particles is governed by the Mathisson--Papapetrou--Dixon equations, which are most commonly applied to massive bodies, but the framework also works in the massless case. Such massless versions naturally arise, for example, in the description of energy centroids of high-frequency wave packets. In this work, we consider massless spinning particles in spacetimes with hidden symmetries and we derive the generalized conservation laws associated with conformal Killing--Yano tensors. We then show that the spin Hall equations, a particular case of the Mathisson--Papapetrou--Dixon equations restricted to massless particles with longitudinal angular momentum, are completely integrable in a large class of type D spacetimes. Additionally, we also show that for massive spinning particles, the generalized Carter constant associated with Killing--Yano tensors is conserved independently of the choice of spin supplementary condition.
\end{abstract}

\maketitle

\section{Introduction}

The motion of extended test objects in general relativity can be systematically described by the Mathisson--Papapetrou--Dixon (MPD) equations, obtained from a multipole expansion of the stress-energy tensor about a reference worldline \cite{mathisson2010republication,Papapetrou,Dixon70a,doi:10.1098/rspa.1970.0191,Dixon74,HarteGrav}. 
These equations are very general, as their derivation only requires the test object to have a conserved stress-energy tensor of sufficiently compact support. 
The MPD framework therefore provides a general description of the dynamics of spinning test particles in curved spacetime and is widely employed in applications ranging from compact-object motion in strong gravitational fields \cite{PhysRevD.111.044009,PhysRevD.105.084031,afshordi2023waveform,Pound2022} to effective semiclassical or high-frequency models for wave packet propagation \cite{rudiger,audretsch,HarteOancea,GSHE_rev}.

Most applications of the MPD equations focus on massive spinning particles, that is, on bodies with timelike linear momentum. 
However, the derivation of the equations does not rely on this restriction, and the formalism in principle applies more generally. 
Therefore, it is also possible to use the MPD framework for the description of massless spinning particles, and several attempts in this direction have been made \cite{souriau1974modele, saturnini1976modele, Mashhoon1975, bailyn1977pole, bailyn1981pole, bini2006massless, semerak2015spinning, Duval2006}. 
More recently, massless versions of the MPD equations have been obtained by considering semiclassical or high-frequency approximations for electromagnetic \cite{GSHE2020,HarteOancea,PhysRevD.109.064020,Frolov2020,PhysRevD.110.064020,SHE_QM1}, linearized gravitational \cite{GSHE_GW,SHE_GW,GSHE_lensing,GSHE_lensing2,Frolov_2024} and Dirac \cite{GSHE_Dirac} test fields in general relativity. 
In this setting, the massless MPD equations describe gravitational spin Hall effects \cite{GSHE_rev}, where the propagation of a wave packet that carries spin angular momentum (as well as other types of angular momentum) is approximately described by its energy centroid. Then, in this point-particle limit, the energy centroid generally deviates from geodesic motion in a spin-dependent way \cite{GSHE_rev}. Gravitational spin Hall effects can also be viewed as a general-relativistic analog of the spin Hall effects that have been experimentally observed in condensed matter physics \cite{SHE_review,originalSHE3,originalSHE4} and in optics \cite{SOI_review,SHEL_experiment,Bliokh2008}. 

Conservation laws and integrability play a central role in understanding the dynamics of test particles and test fields in curved spacetime \cite{Frolov2017}. 
When a spacetime admits Killing vectors, the associated symmetries generate conserved quantities, such as energy and angular momentum, for geodesic motion. 
More broadly, many physically relevant geometries also admit hidden symmetries encoded in more general geometric objects such as Killing--Yano (KY) tensors \cite{yano1952some,bochner1948curvature} and their associated Killing tensors. 
These structures yield additional conservation laws, the most famous being Carter's constant in the Kerr spacetime \cite{Carter:1968cmp,PhysRev.174.1559}, which render the geodesic equations completely integrable. 
Furthermore, these symmetry structures also underpin the separability properties of field equations (including scalar, Maxwell, Dirac, and linearized gravity) on such spacetimes \cite{Carter:1968cmp,PhysRev.174.1559,teukolskyequation,Frolov2017,Andersson:2015xla,Andersson:2014lca}.
In fact, all type D spacetimes possess conformal Killing--Yano (CKY) tensors~\cite{Walker:1970un,PenroseRindler2} which naturally lead to generalizations of these structures \cite{Araneda:2016iwr,Araneda:2018ezs,Aksteiner:2021nec}.

This picture extends to spinning test particles, where the MPD equations are known to admit generalized (spin-dependent) conservation laws associated with Killing vectors and KY tensors \cite{Dixon70a,rudiger1981conserved,rudiger1983conserved,Gibbons1993,Compere:2021kjz,Compere:2023alp,MPD_conservation_2019,HarteSyms}. 
Based on these conserved quantities, the equations of motion have been shown to be completely integrable in specific cases \cite{Ramond_2025}.
An analytic solution for the motion of spinning test particles in the Kerr spacetime, using separability from these constants of motion, has also been constructed \cite{Skoupy:2024uan}.
We stress that for spinning particles in general relativity, the concept of complete integrability is always understood in a perturbative sense \cite{PhysRevD.103.064066,fasano2006analytical}. The MPD equations are always viewed as incorporating small spin-dependent corrections to geodesic motion, where terms of a certain high order in spin are considered to be small and are truncated from the equations. Also, some of the conservation laws (such as the generalized Carter constant) do not hold exactly, but only up to error terms of a certain order in spin.

The existing literature focuses primarily on massive spinning particles. 
In the massless case, while conservation laws associated with Killing \cite{HarteOancea,Duval} and conformal Killing vectors \cite{PhysRevD.111.024034} are established, those derived from hidden symmetries (determined by CKY tensors) are missing. 
Moreover, apart from some particular cases \cite{Frolov_2024a,PhysRevD.111.024034}, general results regarding the integrability of the equations of motion for massless spinning particles remain absent. 

In this work, we focus on the case of massless spinning particles described by the MPD equations in the pole-dipole approximation, where quadrupole and higher-order multipole moments, as well as terms quadratic in spin, are ignored. 
This framework encompasses the spin Hall equations derived in \cite{GSHE2020,GSHE_GW,GSHE_Dirac,HarteOancea} and the spinoptics equations \cite{Frolov2020,Frolov_2024,PhysRevD.110.064020,Frolov_2024a,trh5-4sgq,PhysRevD.111.044001}. 
Under these assumptions, we derive the generalized conservation laws associated with CKY tensors, and we demonstrate that the spin Hall equations satisfy weak complete integrability \cite{SARLET198587} in a perturbative sense \cite{PhysRevD.103.064066,fasano2006analytical}, up to error terms quadratic in spin.
This result holds for a large class of type D spacetimes, including the Pleba\'nski--Demia\'nski \cite{Plebanski:1976gy} and the Ovcharenko--Podolsk\'y \cite{8wkz-th6v} classes of spacetimes. Furthermore, we revisit the massive case to demonstrate that the generalized Carter constant associated with a KY tensor is conserved independently of the choice of spin supplementary condition (SSC).

This paper is structured as follows. 
In \cref{sec:MPD}, we introduce the MPD equations in the pole-dipole approximation, where quadrupole and higher-order multipole moments, as well as terms quadratic in spin, are neglected. 
The worldline equation of motion is determined by introducing a SSC, and we discuss the differences that arise between the massive and massless cases. 
In \cref{sec:geometry}, we review the geometric quantities that can give rise to hidden symmetries and conservation laws for particle dynamics. 
These include CKY tensors, conformal Killing tensors, and conformal Killing vectors. 
We also discuss how the existence of such geometric objects constrains the spacetime geometry through integrability conditions, and we provide some explicit expressions in certain type D spacetimes. 
Our main results are presented in \cref{sec:Cons,sec:integrability}. We first construct the conserved quantities admitted by the MPD equations in the pole-dipole approximation in both the massive and the massless cases. 
The new results here mainly involve the generalized Carter constant: we show that such a constant exists in the massless case for spacetimes that admit a CKY tensor, and that in the massive case its conservation does not depend on the choice of SSC. Based on these results, in \cref{sec:integrability}, we show that the massless version of the MPD equations given by the spin Hall equations is completely integrable in the weak form defined in Ref. \cite{SARLET198587}. 
Finally, we present our conclusions in \cref{sec:conclusions}.

\textit{Notation and conventions.} We work on smooth Lorentzian manifolds $(M, g_{\mu \nu})$, where the metric $g_{\mu \nu}$ has the signature $(-\,+\,+\,+)$. 
The absolute value of the metric determinant is denoted as $g = |\det g_{\mu \nu}|$. 
Phase space is defined as the cotangent bundle $T^*M$, with canonical coordinates $(x, p)$. 
Covariant derivatives of tensor fields on $M$ are denoted by $\nabla_\mu$, and we use a dot to denote the covariant derivative of tensor fields along a worldline. 
We use the Lie bracket notation $[v, T]$ to denote the Lie derivative of a tensor field $T$ along a vector field $v$.

\section{The Mathisson--Papapetrou--Dixon equations} \label{sec:MPD}

In general relativity, the motion of sufficiently compact spinning objects with conserved stress-energy tensors is generally described by the Mathisson--Papapetrou--Dixon equations \cite{mathisson2010republication,Papapetrou,dixon2015new,HarteGrav,HarteReview}
\begin{subequations}
\label{MP}
\begin{align}
    \dot{p}_\mu  &= - \frac{1}{2}  R_{\mu \nu \alpha \beta} \dot{x}^\nu S^{\alpha \beta}, 
    \label{MPp}
    \\
    \dot{S}^{\alpha \beta} &= p^{\alpha} \dot{x}^{\beta} - p^{\beta} \dot{x}^{\alpha},
    \label{MPS}
\end{align}
\end{subequations}
where $p_\mu$ is the linear momentum, $S^{\alpha \beta}$ is the spin tensor that describes the angular momentum carried by the object, and $x^\mu$ is the worldline along which these quantities evolve. 
Depending on the object on which it acts, the dot denotes both the derivative with respect to the affine parameter along the worldline 
\begin{equation}
    \dot{x}^\mu = \frac{d x^\mu (\tau)}{ d \tau},
\end{equation}
as well as the covariant derivative of tensors along the worldline
\begin{equation}
    \dot{p}_\mu = \frac{d p_\mu (\tau)}{d \tau} - \Gamma^{\alpha}_{\mu \beta} p_\alpha \dot{x}^\beta.
\end{equation}

The MPD equations \eqref{MP} hold for all compact objects with conserved stress-energy tensors, provided that the quadrupole and higher-order multipole moments can be ignored \cite{Dixon74, HarteReview}. 
In principle, higher-order multipole moments can be included, but here we focus on the pole-dipole approximation. 
Furthermore, while most of the literature is concerned with the MPD equations with timelike linear momentum $p_\mu$, the derivation of the equations makes no use of this condition. 
Thus, a null linear momentum $p_\mu$ is also possible, leading to a description of massless spinning particles.

The MPD equations \eqref{MP} form an underdetermined system of ordinary differential equations. 
This is because there is no equation that determines the worldline $x^\mu$. 
To close the system and obtain an equation for the worldline, additional information must be provided, usually in the form of a SSC
\begin{equation} \label{eq:SSC_S.t}
    S^{\alpha \beta} t_\beta = 0,    
\end{equation}
where $t_\beta$ is a timelike covector field \cite{Costa2015,HarteOancea}. 
The choice of $t_\beta$ can be physically interpreted as fixing a family of timelike observers with respect to which the center of mass or the energy centroid of the spinning object is defined \cite{Costa2015,Herdeiro2012,HarteOancea}. 

Taking the derivative of the SSC \eqref{eq:SSC_S.t} and using \cref{MPS}, we obtain the worldline equation
\begin{equation} \label{eq:worldline}
    \dot{x}^\alpha = \frac{\dot{x}^\beta t_\beta}{p^\sigma t_\sigma} p^\alpha + \frac{1}{p^\sigma t_\sigma} S^{\alpha \beta} \dot{t}_\beta.
\end{equation}
Furthermore, if we assume that the vectors $\dot{x}^\alpha$ and $p^\alpha$ are causal and future-directed with respect to $t^\alpha$, then the worldline parametrization can be chosen so that 
\begin{equation}
    \frac{\dot{x}^\beta t_\beta}{p^\sigma t_\sigma} = \lambda,
\end{equation}
where $\lambda$ is a positive constant.

The MPD equations \eqref{MP} together with the worldline equation \eqref{eq:worldline} determined by the SSC \eqref{eq:SSC_S.t} hold regardless of the timelike or null character of the linear momentum $p_\mu$.
We discuss in the following some of the possible choices of SSCs for the massive and massless cases separately, as the two cases are significantly different. 
Also, we will generally work in the pole-dipole approximation, ignoring any terms that are quadratic in spin, as well as higher-order multipole moments.

\subsection{Massive spinning particles}

The MPD equations are most often used to describe the dynamics of spinning particles with timelike momentum. 
A common physical scenario in this case is the dynamics of extreme mass-ratio binary black holes \cite{PhysRevD.111.044009,PhysRevD.105.084031,afshordi2023waveform,Pound2022}, but the MPD equations also arise from the semiclassical limit of Dirac wave packets propagating in curved spacetime \cite{audretsch,rudiger,GSHE_Dirac}. 

In this case, a choice often used in the literature to fix the worldline is the Tulczyjew--Dixon SSC \cite{tulczyjew1959motion,dixon1964covariant}
\begin{equation}\label{eq: TD condition}
    S^{\alpha \beta} p_\beta = 0.
\end{equation}
This has the advantage of defining the worldline without introducing any additional vector fields external to the spinning particle. 
Using this SSC, the worldline equation is \cite{HarteGrav,Ehlers1977,rudiger1981conserved}
\begin{equation} \label{eq:dot_x_massive}
    \dot{x}^\mu = \frac{1}{m} p^\mu,
\end{equation}
where the mass parameter is defined as $m = \sqrt{- p_\mu p^\mu}$, and we have ignored the terms quadratic in spin. 
The worldline parametrization is chosen so that 
\begin{equation}
    \lambda = \frac{\dot{x}^\beta p_\beta}{p^\sigma p_\sigma} = \frac{1}{m}.
\end{equation}
This is a proper time parametrization since $\dot{x}^\mu \dot{x}_\mu = -1$.

In general, other choices of SSCs can be used \cite{Costa2015}, and it is also possible to relate worldlines, linear momenta, and angular momenta associated with different SSCs \cite{vines2016canonical}.

\subsection{Massless spinning particles}

The derivation of the MPD equations does not require the assumption that the momentum $p_\mu$ is timelike. 
Null momenta can also be considered, and the MPD equations can describe massless spinning particles. 
However, it has been shown in \cite{HarteOancea} that if the dominant energy condition is assumed to be satisfied, then an object with exactly null momentum $p_a p^a = 0$ must have vanishing angular momentum $S^{\alpha \beta} = 0$. 
In other words, massless spinning objects with exactly null momenta would violate the dominant energy condition. 
Nevertheless, the description of massless spinning particles by the MPD equations is still useful if viewed in an approximate sense, where the momentum is only approximately null up to error terms quadratic in spin (or equivalently, up to error terms quadratic in the wavelength of the wave packet). 
For example, such massless versions of the MPD equations arise when considering the semiclassical or high-frequency dynamics of localized wave packets for certain massless field theories on curved spacetimes, such as electromagnetism \cite{GSHE2020,HarteOancea,PhysRevD.109.064020,Frolov2020,PhysRevD.110.064020,SHE_QM1}, linearized gravity \cite{GSHE_GW,SHE_GW,GSHE_lensing,GSHE_lensing2,Frolov_2024}, and Dirac fields \cite{GSHE_Dirac}.

Thus, we consider massless spinning particles as an approximate model for the dynamics of the energy centroids of high-frequency wave packets. These have approximately null momenta up to error terms quadratic in the angular momentum (or equivalently up to error terms quadratic in the wavelength, since the spin tensor of such wave packets is proportional to the wavelength \cite{HarteOancea})
\begin{equation}\label{eq: null}
    p_\mu p^\mu = \mathcal{O}(S^2).
\end{equation}
This approximate notion of null momenta and the meaning of the error terms $\mathcal{O}(S^2)$ follow, for example, by considering the description of wave packets in terms of semiclassical or high-frequency approximations \cite{GSHE2020,HarteOancea,GSHE_GW,GSHE_Dirac,Frolov2020,Frolov_2024}, where the field is approximated by an ansatz of the form 
\begin{equation}
    \Psi = \Big[\psi_0 + \varepsilon \psi_1 + \mathcal{O}(\varepsilon^2) \Big] e^{i \phi /\varepsilon}.
\end{equation}
Here, $\varepsilon$ is a small expansion parameter related to the wave frequency $\omega$ measured by an observer with $4$-velocity $t^\alpha$ as $\omega \varepsilon = -t^\alpha \nabla_\alpha \phi$. Following Ref. \cite{HarteOancea}, when we go beyond the leading-order geometric optics regime described by null geodesic motion, the linear momentum of the wave packet remains approximately null, $p_\mu p^\mu = \mathcal{O}(\varepsilon^2)$, but the motion of the energy centroid deviates from geodesic motion and becomes spin dependent, leading to spin Hall effects. 
In particular, the spin tensor of such wave packets is proportional to $\varepsilon$, so we have $S^{\alpha \beta} = \mathcal{O}(\varepsilon)$ and we can write $p_\mu p^\mu = \mathcal{O}(\varepsilon^2) = \mathcal{O}(S^2)$.

An important issue in this massless case is that the Tulczyjew--Dixon SSC $S^{\alpha \beta} p_\beta = 0$ cannot be used to fix the worldline \cite{HarteOancea}, \cite[p. 70]{PenroseRindler2}. 
Thus, to fix the definition of the energy centroid and obtain an equation for the worldline, we have to introduce a timelike vector field $t^\alpha$ and use the SSC in \cref{eq:SSC_S.t}. 
As discussed in \cite{HarteOancea}, $t^\alpha$ can be interpreted as being proportional to the $4$-velocity of a family of timelike observers relative to which the energy centroid of a wave packet is defined. 
After differentiating the SSC in \cref{eq:SSC_S.t}, we obtain the worldline equation \eqref{eq:worldline}. 
Additionally, we can fix the worldline parametrization so that $\lambda = 1$ and we obtain 
\begin{equation} \label{eq: xdot rel}
    \dot{x}^\alpha = p^\alpha + \frac{1}{p^\sigma t_\sigma} S^{\alpha \beta} \dot{t}_\beta.
\end{equation}
This equation should be understood in an approximate sense, where the terms quadratic in $S$ are ignored. 
In particular, we could also write $S^{\alpha \beta} \dot{t}_\beta = S^{\alpha \beta} \dot{x}^\mu \nabla_\mu t_\beta = S^{\alpha \beta} p^\mu \nabla_\mu t_\beta + \mathcal{O}(S^2)$. 

Thus, \cref{MP,eq: xdot rel} provide a description of massless spinning particles with approximately null linear momentum. 
However, we can also introduce additional constraints on the type of angular momentum carried by such objects. 

The spin Hall equations introduced in \cite{GSHE2020,GSHE_GW,HarteOancea} are recovered if we also impose the physical constraint $S^{\alpha \beta} p_\beta = 0$. 
Note that here this relation is not used as a SSC, but as an additional physical constraint on the type of angular momentum. 
More precisely, this is a restriction to the case of wave packets with longitudinal angular momentum \cite{HarteOancea}. 
Thus, imposing simultaneously the SSC $S^{\alpha \beta} t_\beta = 0$ and the constraint $S^{\alpha \beta} p_\beta = 0$ fixes the spin tensor up to its magnitude. 
Using an adapted tetrad $\{p_\alpha, t_\alpha, m_\alpha, \bar{m}_\alpha \}$, where $p_\alpha t^\alpha = - \varepsilon \omega$, $t_\alpha t^\alpha = -1$, $m_\alpha \bar{m}^\alpha = 1$ and all other contractions are zero, we can write
\begin{align} \label{eq:S_SHE}
    S^{\alpha\beta} &= 2i\varepsilon s \bar{m}^{[\alpha} m^{\beta]} = \frac{\varepsilon s }{p\cdot t} \epsilon^{\alpha\beta\gamma\delta} p_\gamma t_\delta .
\end{align}
The magnitude of the spin tensor is determined by fixing the constants $\varepsilon s$, where  $\varepsilon$ is a small expansion parameter related to the wavelength of the wave packet, and the dimensionless constant $s$ describes the amount of longitudinal angular momentum carried by the wave packet.  
For example, if we restrict ourselves to wave packets that only carry spin angular momentum, we have $s = \pm 1$ for circularly polarized electromagnetic wave packets \cite{GSHE2020}, and $s = \pm 2$ for linearized gravitational wave packets of circular polarization \cite{GSHE_GW}. Higher values of $s$ are, in principle, possible if we consider wave packets that carry intrinsic orbital angular momentum \cite{BLIOKH20151,HarteOancea}.

With the spin tensor determined as above, the spin Hall equations are 
\begin{subequations} \label{eq:SHE}
\begin{align}
    \dot{x}^\alpha &= p^\alpha + \frac{1}{p^\sigma t_\sigma} S^{\alpha \beta} \dot{t}_\beta, \label{eq:SHE_x}\\
    \dot{p}_\mu  &= - \frac{1}{2}  R_{\mu \nu \alpha \beta} \dot{x}^\nu S^{\alpha \beta}.
\end{align}
\end{subequations}
These equations have been derived using semiclassical or high-frequency approximations for electromagnetic \cite{GSHE2020,HarteOancea}, linearized gravitational \cite{GSHE_GW}, and massless Dirac fields \cite{GSHE_Dirac}. 
Furthermore, a particular case of these equations was also obtained using similar methods in \cite{Frolov2020,Frolov_2024,PhysRevD.110.064020}. 
In this case, the timelike vector field $t^\alpha$ is chosen to be parallel transported along the worldline $x^\mu$ up to the leading order in spin, so that $\dot{t}^\alpha = \mathcal{O}(S)$ and the second term in \cref{eq:SHE_x} becomes $\mathcal{O}(S^2)$ and is ignored.

\section{Conformal Killing--Yano tensors and hidden symmetries} \label{sec:geometry}

The analysis of the spin Hall equations \eqref{eq:SHE}, as for all test particle and field equations, typically takes place on a background spacetime where one can make use of the isometries and Killing vectors (and related constants of motion) of the spacetime to integrate the equations. 
We seek to generalize this procedure to type D spacetimes that possess hidden symmetries, namely Killing and (conformal) Killing--Yano tensors \cite{yano1952some,bochner1948curvature}. 
As will be discussed in the next section, these objects are called hidden symmetries because the resulting constants of motion for test particles typically depend on the phase space variables, including both the worldline and the momentum of the particle.

For this purpose, it is necessary to introduce the following catalog of objects. We start with a rank-$p$ CKY tensor $h$. In $d$ dimensions this is a $p$-form whose covariant derivative can be written as \cite[Eq. (2.50)]{Frolov2017}
\begin{equation}\label{eq: CKY}
    \nabla_{\beta}h_{\alpha_1\dots \alpha_p}=\nabla_{[\beta} h_{\alpha_1\dots \alpha_p]}+\frac{p}{d-p+1}g_{\beta[\alpha_1}\nabla_{|\sigma|}h^{\sigma}{}_{\alpha_2\dots\alpha_p]}.
\end{equation}
When the first term vanishes, $h$ is closed and is hence called a closed CKY tensor.
This is actually an extremely restrictive condition that uniquely fixes the spacetime to be the off-shell Carter metric \cite{Houri:2007xz,Ferrando:2008nw,Krtous:2008tb} introduced below in \cref{sec: Carter}.
When both terms vanish, $h$ is a Killing--Yano (KY) tensor that clearly satisfies the more standard defining equation
\begin{equation}
    \nabla_{(\beta}h_{\alpha_1)\dots \alpha_p}=0.
\end{equation}

Any two rank-$p$ CKY tensors automatically generate a rank-2 conformal Killing tensor~\cite{Frolov2017},
\begin{equation}
   \K^{\alpha\beta}= [h_1]^{(\alpha}{}_{\gamma_1\dots\gamma_{p-1}} [h_2]^{\beta)\gamma_1\dots\gamma_{p-1}},
\end{equation}
which satisfies
\begin{subequations}\label{eq:def_CKV}
\begin{align}
    \nabla_{(\alpha}\K_{\beta\gamma)} &= g_{(\alpha\beta}\tilde{\K}_{\gamma)}, \\
    \tilde{\K}_\alpha &= \frac{1}{d+2}(2\nabla_\rho \K^\rho{}_\alpha+\nabla_\alpha \K^\rho{}_\rho).
\end{align}
\end{subequations}
In the case where $h_1$ and $h_2$ are KY tensors [namely the second term in \cref{eq: CKY} vanishes], then $\K^{\alpha\beta}:=K^{\alpha\beta}$ is a Killing tensor which satisfies
\begin{equation}
     \nabla_{(\alpha}K_{\beta\gamma)}=0.
\end{equation}
The generalization to a rank-$p$ (conformal) Killing tensor $\K_{(\alpha_1\dots\alpha_p)}=\K_{\alpha_1\dots\alpha_p}$ is immediate,
\begin{equation}
    \nabla_{(\alpha_1}\K_{\alpha_2\dots\alpha_{p+1})}=g_{(\alpha_1\alpha_2}\tilde{\K}_{\alpha_3\dots\alpha_{p+1})}.
\end{equation}
Note that a rank-1 (conformal) Killing tensor $\kappa^\beta$, is a (conformal) Killing vector and satisfies
\begin{equation} \label{eq:def_CKV1}
    \nabla_{(\alpha}\kappa_{\beta)} = \frac{1}{d} g_{\alpha\beta}\ \nabla_\gamma \kappa^\gamma,.
\end{equation}

Finally, a rank-$p$ CKY tensor $h_{\alpha_1\dots\alpha_p}$ automatically generates a rank-$(d-p)$ CKY tensor by Hodge duality:
\begin{equation}
f_{\alpha_1\dots\alpha_{d-p}}= [\star h]_{\alpha_1\dots\alpha_{d-p}}=\frac{1}{p!}\epsilon_{\alpha_1\dots\alpha_{d-p}}{}^{\beta_1\dots\beta_p}h_{\beta_1\dots\beta_p}.
\end{equation}
In addition, if $h_{\alpha_1\dots\alpha_p}$ is a closed CKY tensor, then $f=\star h$ is a KY tensor.
For further discussion of these objects, we refer the reader to \cite{Frolov2017} and the references therein. In addition, for $d=4$ CKY tensors are equivalent to Killing spinors \cite{Walker:1970un,PenroseRindler2,Andersson:2015xla}.

\subsection{Integrability conditions}\label{sec: int cons}

Although all of these equations contain only first derivatives of the objects in question, they actually impose conditions on the second derivatives of these objects because they are overdetermined. 
This puts severe constraints on the geometry. In particular, certain integrability conditions that involve the curvature tensors must be satisfied \cite{Frolov2017}.
For example, consider a conformal Killing vector $\kappa^\alpha$ satisfying \cref{eq:def_CKV1}. Then, a straightforward calculation (see, for example, \cite[App. C]{Wald}) shows that the following integrability condition must be satisfied:
\begin{equation}\label{CKV: int con}
    \nabla_\alpha\nabla_\beta \kappa_\gamma = R\indices{_{\gamma \beta \alpha \delta}} \kappa^\delta + \left(g_{\beta \gamma}\nabla_\alpha + g_{\gamma \alpha} \nabla_\beta - g_{\alpha \beta} \nabla_\gamma \right) \sigma,
\end{equation}
where $\sigma = \tfrac{1}{d} \nabla^\mu \kappa_\mu$
The corresponding integrability condition for a Killing vector is obtained by setting $\sigma = 0$ in the above equation.

This kind of condition extends to the catalog mentioned above. In particular, relevant for our work, given a rank-$2$ CKY $h_{\alpha\beta}$, we have
\begin{align}\label{eq: Int cons}
    2 \nabla_{[\mu} \nabla_{\nu]} h_{\rho \alpha} &= R_{\mu \nu \alpha \sigma} h\indices{_\rho^\sigma} + R_{\mu \nu \rho \sigma} h\indices{_\alpha^\sigma} \nonumber \\
    &= 2\nabla_\mu \xi_{[\alpha} g_{\rho]\nu} - 2 \nabla_\nu \xi_{[\alpha} g_{\rho] \mu}
    +\frac{2}{3}\nabla_{[\mu}dh_{\nu]\rho\alpha} ,
\end{align}
where $dh_{\nu \rho \alpha} = 3 \nabla_{[\nu} h_{\rho \alpha]}$. Moreover~\cite{Batista:2014fpa},
\begin{align}\label{eq: grad grad h}
    2\nabla_{\alpha}\nabla_{\beta}h_{\gamma\delta}
    &=-3(R^\epsilon{}_{\alpha[\beta\gamma}h_{\delta]\epsilon}-2g_{\alpha[\beta}\nabla_{\gamma}\xi_{\delta]})
    \nonumber\\
    &\quad+2(g_{\beta\gamma}\nabla_{\alpha\delta}-g_{\beta\delta}\nabla_{\alpha}\xi_{\gamma}),
\end{align}
and this gives
\begin{align}\label{eq: grad dh}
    \frac{2}{3}\nabla_{\mu}dh_{\nu\rho\alpha} &= R_{\mu \nu [\rho}{}^{\sigma} h_{\alpha]\sigma} + R_{\alpha \rho [\mu}{}^{\sigma} h\indices{_{\nu]}_\sigma}
    \nonumber\\
    &\quad-2(\nabla_\mu \xi_{[\alpha} g_{\rho]\nu} - \nabla_\nu \xi_{[\alpha} g_{\rho] \mu}).
\end{align}
When $dh=0$, we can use the closed CKY integrability condition in \cite[Eq. (C.26)]{Frolov2017} to show that \cref{eq: grad dh} is identically zero, as required.
Next, combining \cref{eq: grad dh,eq: Int cons} we have
\begin{align}
     2\nabla_{[\mu} \nabla_{\nu]} h_{\rho \alpha} &=R_{\mu \nu [\rho}{}^{\sigma} h_{\alpha]\sigma} + R_{\alpha \rho [\mu}{}^{\sigma} h\indices{_{\nu]}_\sigma}
     .
\end{align}

Finally, an identity can be obtained that involves only the CKY tensor and the curvature~\cite{Batista:2014fpa},
\begin{align}\label{eq: Curv Condtions}
    I_{\alpha\beta\gamma\delta}&=h_{\alpha \epsilon}R\indices{^{\epsilon}_{\delta\beta\gamma}}+h_{\beta\epsilon}R\indices{^{\epsilon}_{\gamma\alpha\delta}}+h_{\gamma\epsilon}R\indices{^{\epsilon}_{\beta\delta\alpha}}+h_{\delta\epsilon}R\indices{^{\epsilon}_{\alpha \gamma\beta}}
    \nonumber\\
    &\quad+2(g_{\alpha\gamma}P_{\beta\delta}+g_{\beta\delta}P_{\alpha\gamma}-g_{\alpha\beta}P_{\gamma\delta}-g_{\gamma\delta}P_{\alpha\beta})
    \nonumber\\
    &=0,
\end{align}
where we have introduced the tensor
\begin{equation}\label{eq: h Ricc rel}
  P_{\alpha\beta}=\frac{1}{d-2}h_{(\alpha|\gamma|}R^\gamma{}_{\beta)}.
\end{equation}
See also Refs. \cite{tachibana1969integrability,Houri:2017tlk,Houri:2014hma} for more details on the integrability conditions for the other objects in our catalog.

\subsection{\texorpdfstring{$d = 4$ spacetimes admitting hidden symmetries}{d = 4 spacetimes admitting hidden symmetries}}

Specializing now to four dimensions, as mentioned above, Penrose and Walker~\cite{Walker:1970un,Andersson:2015xla} have shown that all Petrov type D spacetimes admit a Killing spinor that is equivalent to a CKY tensor.
Thus, these objects exist for physically relevant spacetimes such as Kerr spacetimes~\cite{Kerr:1963ud}. 
Hence, we assume going forward that the spacetime admits a rank-$2$ CKY $2$-form,
\begin{equation} \label{eq:h_def}
    \nabla_\mu h_{\alpha \beta} = \nabla_{[\mu} h_{\alpha\beta]}+g_{\mu \alpha} \xi_\beta - g_{\mu \beta} \xi_\alpha,
\end{equation}
where $\xi_\beta = \frac{1}{3} \nabla^\alpha h_{\alpha \beta}$. Moreover, Jezierski and {\L}ukasik  \cite{Jezierski_2006} have shown that such a vector satisfies
\begin{equation}\label{eq: KV obstr}
    \nabla_{(\alpha}\xi_{\beta)}=P_{\alpha\beta},
\end{equation}
where $P_{\alpha\beta}$ was defined in \cref{eq: h Ricc rel}.
Thus, for Einstein manifolds (defined by the fact that the Ricci tensor is proportional to the metric) in arbitrary dimensions, either $\xi$ vanishes (in which case $h$ is a KY tensor), or $\xi$ is a Killing vector. 
Notice that when $P_{\alpha\beta}$ vanishes, the second line of \cref{eq: Curv Condtions} is also zero.

The dual CKY tensor $f_{\alpha\beta}=[\star h]_{\alpha\beta}$ satisfies the same kind of equation 
\begin{equation} \label{eq: dual CKY}
    \nabla_\mu f_{\alpha \beta} = \nabla_{[\mu} f_{\alpha\beta]}+g_{\mu \alpha} \zeta_\beta - g_{\mu \beta} \zeta_\alpha,
\end{equation}
where $\zeta_\beta=\frac{1}{3}\nabla^\alpha f_{\alpha\beta}$.  Again,  for Einstein spacetimes, $\zeta$ either vanishes or is a (possibly distinct) Killing vector.

The resulting conformal Killing tensor $\K_{\alpha \beta}$ can be expressed in terms of the CKY tensor as
\begin{equation} \label{eq:K_def}
    \K_{\alpha \beta} = h_{\alpha \mu} h\indices{^\mu_\beta}.
\end{equation}
If $h$ is closed, then a Killing tensor $K_{\mu\nu}$ can be obtained by squaring the KY tensor, $f$, instead. 
In terms of $h$, this can be written explicitly in four dimensions as
\begin{equation}\label{eq: KT h}
    {K}_{\mu\nu}=\K_{\mu\nu}+\frac 1 2 h_{\alpha\beta} h^{\alpha\beta} g_{\mu \nu},
\end{equation}
where $\K_{\mu\nu}$ is the conformal Killing tensor from \cref{eq:K_def}.

\subsubsection{Carter spacetimes}\label{sec: Carter}

The archetypal example of such a spacetime is the previously mentioned Kerr spacetime which we present in Carter coordinates $(\tau, r, y, \psi)$~\cite{Carter:1968cmp},
\begin{align}\label{eq: Carter met}
    g^{\text{C}}&=-\frac{\Delta_r}{\Sigma}(d\tau+y^2 d\psi)^2+\frac{\Delta_y}{\Sigma}(d\tau-r^2 d\psi)^2\nonumber\\
       &\qquad+\frac{\Sigma}{\Delta_r}dr^2+\frac{\Sigma}{\Delta_y}dy^2, 
\end{align}
where 
\begin{equation}\label{Sigmadef}
\Sigma=\sqrt{-\det (g) }=r^2+y^2,
\end{equation}
and the metric functions $\Delta_r$ and $\Delta_y$ each depend on only one variable,
\begin{equation} \label{generalDelta}
\Delta_y=\Delta_y(y), \qquad \Delta_r=\Delta_r(r).
\end{equation}
When these are arbitrary, this is known as the {\em off-shell canonical} metric, for which most of the symmetry properties remain valid \cite{Frolov2017}.  

Imposing the Einstein--Maxwell equations with the cosmological constant $\Lambda$, fixes the metric functions to take the following specific form: 
\begin{subequations} \label{Delr}
\begin{align}
\Delta_r &=(r^2+a^2)(1-\Lambda r^2/3)-2Mr+q_e^2+q_m^2 ,\\
\Delta_y &=(a^2-y^2)(1+\Lambda y^2/3)+2Ny .
\end{align}
\end{subequations}
Here, $M, a$ are the mass and angular momentum per unit mass, $q_e$ and $q_m$ are the electric and magnetic charges, and $N$ is the Newman--Unti--Tamburino (NUT)~\cite{NewmanEtal:1963} parameter.
In addition, the electromagnetic potential is given by
\begin{equation}\label{ACarter}
A^{\text{C}}=-\frac{1}{r^2+y^2}\Bigl[q_m y(d\tau+r^2 d \psi)-q_e r(d \tau-y^2 d \psi)\Bigr].
\end{equation}
This generalization of the Kerr metric is known as Kerr--NUT-(A)dS.
This metric obviously has two explicit symmetries corresponding to the  Killing vectors
\begin{equation}
    k=\partial_\tau,\qquad \ell=\partial_\psi,
\end{equation}
where $k$ is the generator of time translations and $\ell$ is related to axial rotation.
Moreover, the hidden symmetries arise from a \emph{closed} CKY whose explicit form is
\begin{equation}\label{eq: h Carter}
    h^\text{C}= y dy\wedge(d\tau-r^2 d\psi)-r dr\wedge (d\tau+y^2 d\psi).
\end{equation}
Remarkably, even off shell, we have the explicit relation between the divergence of the closed CKY $h$ (given by \cref{eq:h_def}) and the generator of time translations $k$,
\begin{equation}
    \xi_\alpha=\frac{1}{3}\nabla^\beta h^\text{C}_{\beta\alpha}=k_\alpha.
\end{equation}
Thus, $P_{\alpha\beta}=0$ off shell, going beyond the statements in~\cite{Jezierski_2006}. 
Note that while \cref{eq: dual CKY} holds, as $h$ is closed $f$ is KY and so $\zeta=0$. 
Moreover, in this case ${K}$ is a Killing tensor that generates the axial Killing vector%
\footnote{Note that \cref{eq: Killing tower} does not hold for the Schwarzschild metric where the right-hand side gives zero ~\cite{hughston1973symmetries}. However, the limit $a\to0$ in the Carter metric is a little subtle due to the choice of coordinates \cite[Sec. 3.6]{Frolov2017}.}
\begin{equation}\label{eq: Killing tower}
    \ell^\alpha={K}^\alpha{}_{\beta}\xi^\beta.
\end{equation}
All of these equations are valid, independently of the Einstein--Maxwell equations, that is, for arbitrary metric functions $\Delta_r$ and $\Delta_y$.
Thus, $h$ is called the principal tensor from which all other symmetries descend. In higher dimensions, this leads to a rich structure known as the Killing tower  \cite{Frolov2017}.

\subsubsection{Pleba\'nski--Demia\'nski spacetimes}
\label{sec:PD_spacetime}

Another relevant example of a type D metric is the more general Pleba\'nski--Demia\'nski family \cite{Plebanski:1976gy}. 
These are again solutions to the Einstein--Maxwell equations. 
Off shell it is conformal to the Carter spacetime and the metric and vector potential are given by
\begin{subequations} \label{eq: PD met}
\begin{align}
    g^{\text{PD}} &= \frac{1}{\Omega(r,y)^{2}} g^{\text{C}},\\
    A^{\text{PD}} &= A^{\text{C}},
\end{align}    
\end{subequations}
with  $g^{\text{C}}$ and $A^{\text{C}}$ given in \cref{eq: Carter met,ACarter}.

On imposing the Einstein--Maxwell equations, the metric functions now take the form
\begin{subequations}
\begin{align}
\Omega^2&={(1-yr)^2},\\
\Delta_r &= \gamma + q_e^2 + q_m^2 - 2 M r + \delta r^2 - 2 N r^3 - (\gamma + \Lambda/3) r^4, \\
\Delta_y &= \gamma + 2 N y - \delta y^2 + 2 M y^3 - (\gamma + q_e^2 + q_m^2 + \Lambda/3 ) y^4 .
\end{align}    
\end{subequations}
Here, $\Lambda$ stands for the cosmological constant, $q_e$ $q_m$ are the electric and magnetic charges, and the four parameters $\{\gamma, \delta, M, N\}$ are related to mass, rotation, acceleration and NUT charge, as discussed in  \cite{Griffiths:2005qp,PhysRevD.111.024038,Ovcharenko:2025fxg}.
When the acceleration vanishes, the Pleba\'nski--Demia\'nski metric reduces to Kerr--NUT-(A)dS, although the parameters need to be rescaled as shown in \cite{Griffiths:2005qp,PhysRevD.111.024038,Ovcharenko:2025fxg}.

Clearly, the metric in \cref{eq: PD met} still has the explicit symmetries $k$ and $\ell$.
Although, this time the hidden symmetries are slightly weaker, we still have a CKY tensor given by~\cite{Kubiznak:2007kh}
\begin{equation} \label{eq:CKY_PD}
    h^{\text{PD}}=\frac{1}{\Omega^3}h^{\text{C}}.
\end{equation}
However, when the acceleration is not zero,%
\footnote{Again, in these coordinates setting the acceleration to zero is subtle, as discussed in \cite{Griffiths:2005qp,PhysRevD.111.024038,Ovcharenko:2025fxg}.}
it is no longer closed, so we only have conformal Killing tensors.  
The divergence still yields the Killing vector $\xi_\beta=\frac{1}{3}\nabla_\alpha h^{\alpha}{}_\beta=k_\beta$, but \cref{eq: Killing tower} is no longer valid. 
On the other hand, the second Killing vector, $\ell=\partial_\psi$, is now generated by the dual CKY tensor $f_{\alpha\beta}=[\star h]_{\alpha\beta}$, i.e. \cref{eq: dual CKY} still holds with $\zeta=\ell=\partial_\psi$. 

It is not unexpected that for both the Carter and the Pleba\'nski--Demia\'nski spacetimes, we recover Killing vectors from the (closed) CKY tensors.
Although the spacetimes are no longer vacuum and we have $R_{\alpha\beta}\not\propto\Lambda g_{\alpha\beta}$ due to electromagnetic charges, the electromagnetic field is aligned with the principal null directions (which are the eigenvectors of $h$) and therefore $P_{\alpha\beta}=0$ in \cref{eq: KV obstr}.

On the other hand, there is a very recent generalization of these spacetimes by Ovcharenko and Podolsk\'y~\cite{8wkz-th6v,rfgv-ybz5}, which is still type D and conformally related to the Carter metric \eqref{eq: Carter met}, but has a nonaligned electromagnetic field.
Thus, while it has two dual CKYs and two explicit Killing vectors $k$ and $\ell$, the hidden symmetries do not generate the explicit ones~\cite{Gray:2025lwy}. 
That is, the divergences of the CKYs $\xi^\beta=\frac{1}{3}\nabla_\alpha h^{\alpha\beta}$ and  $\zeta^\beta=\frac{1}{3}\nabla_\alpha f^{\alpha\beta}$ [based on \cref{eq:h_def,eq: dual CKY}] are not Killing vectors.

Importantly, in all these spacetimes (and in type D spacetimes generically), the Killing vectors Lie commute with each other and with the (conformal) Killing tensor, $\K$~\cite{Frolov2017},
\begin{equation}\label{eq: KVT SN comm}
    [k,\ell]=0,\qquad [k,\K]=0=[\ell,\K].
\end{equation}
These equations hold as well for the Killing tensor $K$, when it exists.
Finally, since the two Killing vectors commute, one can use Cartan's identity, the fact that $h$ and $\Omega$ are independent of the Killing coordinates, and the defining equation \cref{eq:h_def} to show that the Lie derivative of $h$ along each Killing direction vanishes
\begin{equation}\label{eq: h Lie deriv}
    [k, h] = 0 = [\ell, h].
\end{equation}
These last identities, \cref{eq: KVT SN comm,eq: h Lie deriv}, will prove crucial to the integrability of the spin Hall equations.

\section{Conserved quantities}
\label{sec:Cons}

In this section, we present the conserved quantities for spinning particles associated with the geometric quantities and spacetimes discussed in the previous section. 
We consider the MPD equations as discussed above, working up to linear order in spin and ignoring terms quadratic in spin, as well as quadrupole and higher-order multipole moments. 

We start by presenting the conservation laws that hold for all choices of SSC and for both timelike and null momenta. 
These quantities rely on the existence of Killing vectors and KY tensors. 
In contrast to previous claims in the literature \cite{rudiger1981conserved,rudiger1983conserved}, we show that the existence of a generalized Carter's constant defined in terms of a KY tensor is independent of the SSC.

Next, we present the conservation laws that only hold under additional assumptions, such as $S^{\alpha \beta} p_\beta = 0$ or that the linear momentum is approximately null. 
For the null case, the geometric requirements are the existence of conformal Killing vectors and CKY tensors.

\subsection{General conservation laws}

Consider the MPD equations, together with the SSC in \cref{eq:SSC_S.t}, where $t^\alpha$ is an arbitrary timelike vector field. 
The complete set of equations is
\begin{subequations} \label{eq:MPD_cons}
\begin{align}
    \dot{x}^\alpha &= \lambda p^\alpha + \frac{1}{p^\sigma t_\sigma} S^{\alpha \beta} \dot{t}_\beta, \label{eq:dot_x}\\
    \dot{p}_\mu  &= - \frac{1}{2}  R_{\mu \nu \alpha \beta} \dot{x}^\nu S^{\alpha \beta}, \label{eq:dot_p} \\
    \dot{S}^{\alpha \beta} &= p^{\alpha} \dot{x}^{\beta} - p^{\beta} \dot{x}^{\alpha},
\end{align}
\end{subequations}
where $\lambda$ is a positive constant fixed by the choice of worldline parametrization. The linear momentum $p_\mu$ can be timelike or null.

This system of equations admits the following conserved quantities:
\begin{subequations} \label{eq:cons_general}
\begin{align}
    H &= \frac{1}{2} g^{\mu \nu} p_\mu p_\nu, \\
    C &= k^\mu p_\mu + \frac{1}{2} S^{\mu \nu} \nabla_\mu k_\nu, \label{eq: C def}\\
    D &= K^{\mu \nu} p_\mu p_\nu + L_{\alpha \beta \mu} S^{\alpha \beta} p^\mu,\label{eq: D def}
\end{align}
\end{subequations}
where $k^\mu$ is a Killing vector. 
Conservation of $D$ requires the existence of a closed CKY tensor $h_{\alpha \beta}$, such that $\xi_\beta = \tfrac{1}{3}\nabla^\alpha h_{\alpha \beta}$ is a Killing vector, $f_{\alpha \beta} = [\star h]_{\alpha \beta}$ is a KY tensor and $K_{\alpha \beta} = f\indices{_\alpha^\mu} f_{\mu \beta}$ is a Killing tensor.
As mentioned above, this restricts the conservation of such a $D$ to off-shell Carter spacetimes.
The tensor $L_{\alpha \beta \mu}$ is defined in terms of these geometric quantities as
\begin{align} \label{eq:L_def0}
    L_{\alpha \beta \mu} &= -2 \left( h_{\alpha\beta} \xi_\mu + \xi_{[\alpha}h_{\beta]\mu}  +\xi^\sigma h_{\sigma [\alpha} g_{\beta]\mu}\right) \nonumber \\
    &= 2 \left(\epsilon\indices{_{\sigma \alpha \rho [\beta}} f\indices{^\rho_{\mu]}}-\epsilon\indices{_{\sigma \beta \rho [\alpha}} f\indices{^\rho_{\mu]}}
\right)\xi^\sigma .
\end{align}
Notice that the last term in the first equality of \eqref{eq:L_def0} vanishes from the expression $L_{\mu\nu\alpha}S^{\mu\nu}p^\alpha$ when the Tulczyjew--Dixon SSC, \cref{eq: TD condition}, is imposed. 
The previously known $L_{\alpha\beta\mu}$ from the literature is simply the second line.

The conservation of $H$, up to error terms of $\mathcal{O}(S^2)$, follows as
\begin{align}
    \dot{H} = \dot{p}_\mu p^\mu = - \frac{1}{2}  R_{\mu \nu \alpha \beta} p^\mu \dot{x}^\nu S^{\alpha \beta} = \mathcal{O}(S^2),
\end{align}
where we used \cref{eq:dot_x,eq:dot_p}. 

The conservation of $C$ requires the existence of a Killing vector field $k^\alpha$ and holds exactly. 
We have
\begin{align} \label{eq:dot_C}
    \dot{C} &= p_\mu \dot{x}^\nu \nabla_\nu k^\mu + k^\mu \dot{p}_\mu + \frac{1}{2} \dot{S}^{\mu \nu} \nabla_\mu k_\nu + \frac{1}{2} S^{\mu \nu} \dot{x}^\alpha \nabla_\alpha \nabla_\mu k_\nu \nonumber\\
    &= \lambda p^\mu p^\nu \nabla_\nu k_\mu + \frac{1}{p \cdot t} p^\mu S^{\nu \alpha} \dot{t}_\alpha \nabla_\nu k_\mu - \frac{1}{2} R_{\mu \nu \alpha \beta} k^\mu \dot{x}^\nu S^{\alpha \beta} \nonumber\\
    &\quad + \frac{1}{p \cdot t} p^{[\mu} S^{\nu] \alpha} \dot{t}_\alpha \nabla_\mu k_\nu + \frac{1}{2} R_{\rho \alpha \mu \nu} k^\rho \dot{x}^\alpha S^{\mu \nu} \nonumber\\
    &= 0.
\end{align}
The first equality follows from the definition of $C$, and the second equality is obtained by using \cref{eq:MPD_cons} and the integrability condition \eqref{CKV: int con} for the Killing vector. 
The final equality follows from the defining property of the Killing vector, $\nabla_\mu k_\nu = \nabla_{[\mu} k_{\nu]}$. 
Note that $C$ is conserved exactly and not only up to some order in $S$. 
In fact, this quantity is exactly conserved even if higher-order multipole moments are included in the MPD equations \cite{HarteReview}.

The conservation of $D$ is usually discussed in the literature under the assumption that $S^{\alpha \beta} p_\beta = 0$ \cite{rudiger1981conserved,rudiger1983conserved,Compere:2021kjz,Ramond_2025,Compere:2023alp}. 
However, we show here that this condition is not required, at least for conservation up to error terms of $\mathcal{O}(S^2)$. 
Taking the derivative of $D$, we obtain
\begin{align}\label{eq: D conservation}
    &\dot{D} = \dot{K}^{\mu \nu} p_\mu p_\nu + 2 K^{\mu \nu} \dot{p}_\mu p_\nu + \dot{L}_{\alpha \beta \mu} S^{\alpha \beta} p^\mu \nonumber\\
    &\qquad+ L_{\alpha \beta \mu} \dot{S}^{\alpha \beta} p^\mu + L_{\alpha \beta \mu} S^{\alpha \beta} \dot{p}^\mu \nonumber\\
    &= \lambda p^\mu p^\nu p^\sigma \nabla_\sigma K_{\mu \nu} +  \lambda p^\mu p^\nu S^{\alpha \beta} \left( \nabla_\nu L_{\alpha \beta \mu} - K\indices{^\rho_\mu} R_{\rho \nu \alpha \beta} \right) \nonumber\\
    &\quad + \frac{1}{p \cdot t} p^\mu p^\nu S^{\alpha \beta} \dot{t}_\beta \left( \nabla_\alpha K_{\mu \nu} - 2 L_{\alpha \nu \mu} \right) + \mathcal{O}(S^2). 
\end{align}
The first equality follows from the definition of $D$, and the second equality is obtained by using \cref{eq:MPD_cons}. 
Notice that \cref{eq: D conservation} implies three separate equations when $\dot{t}^\alpha$ has a zeroth order term in $S$, such that $\dot{t}^\alpha=\dot{t}_0^\alpha+{\cal O}(S)$. 
Namely, we have
\begin{subequations} \label{eq: D_cons}
\begin{align}
     p^{\alpha} p^\beta p^\gamma \nabla_\gamma K_{\alpha \beta}&={\cal O}(S^2),
     \label{eq: P cubed}
     \\
      S^{\mu \nu} p^\alpha p^\beta (\nabla_\beta L_{\mu \nu \alpha} - R_{\sigma \alpha \mu \nu} K\indices{^\sigma_\beta})&={\cal O}(S^2),
      \label{eq: P2 S}\\
  S^{\mu \nu} p^\alpha p^\beta \dot{t}_\nu ( \nabla_\mu K_{\alpha \beta} - 2 L_{\mu \alpha \beta})&={\cal O}(S^2).
  \label{eq: P2 S t}
\end{align}
\end{subequations}
However, when $t^\alpha=p^\alpha$, using \cref{eq:dot_p} we find that the last equation \eqref{eq: P2 S t} is ${\cal O}(S^2)$ and therefore it is not required to be satisfied at this order in perturbation.

Clearly \cref{eq: P cubed} is satisfied exactly when $K^{\alpha\beta}$ is a Killing tensor. 
In order to solve \cref{eq: P2 S} one can make a general argument in terms  of which possible tensors can comprise $L_{\alpha\beta\mu}$, similar to the original works of R\"udiger \cite{rudiger1981conserved,rudiger1983conserved} or more recently \cite{Compere:2021kjz}. 
However, it is simpler to assume the existence of a closed CKY tensor $h_{\mu\nu}$ [\cref{eq:h_def} with $\nabla_{[\mu}h_{\nu\rho]}=0$] and then construct possible combinations of $h$ and its Killing vector derivative $\xi_\mu=\frac{1}{3}\nabla^\nu h_{\nu\mu}$, as in \cite{Gray:2024kad}. 
Following either approach, one is led directly to \cref{eq:L_def0}, which, using the integrability conditions, can show that \cref{eq: Curv Condtions} solves \eqref{eq: P2 S} \emph{without} imposing the particular Tulczyjew--Dixon SSC.
Finally, \cref{eq: P2 S t} is also solved for the very same $L_{\alpha\beta\mu}$ in \cref{eq:L_def0}. 
This time it is a straightforward calculation when the Killing tensor is expressed in terms of $h_{\alpha\beta}$ through \cref{eq:K_def,eq: KT h}.

The conservation of these quantities has been known in the massive case \cite{rudiger1981conserved,rudiger1983conserved,Compere:2021kjz,Ramond_2025,Compere:2023alp}, although restricted by the assumption that $S^{\alpha \beta} p_\beta = 0$ for $D$. 
For the massless case, the conservation of $H$ and $C$ has been discussed in \cite{HarteOancea} for the spin Hall equations \eqref{eq:SHE}, but the conservation of $D$ is a new result.

\subsection{\texorpdfstring{$S^{\alpha \beta} p_\beta = \mathcal{O}(S^2)$ and $\dot{t}_\alpha = \mathcal{O}(S)$}{Sab pb = O(S2) and ta = O(S)}}

An additional conserved quantity can be obtained if we assume $S^{\alpha \beta} p_\beta = \mathcal{O}(S^2)$ and $\dot{t}_\alpha = \mathcal{O}(S)$. 
These conditions are satisfied when the Tulczyjew--Dixon SSC is used for the description of massive spinning particles as in \cref{eq:dot_x_massive}, but also for massless spinning particles described by the spin Hall equations \eqref{eq:SHE} when the timelike covector field $t_\alpha$ is chosen to be parallel transported along the worldline, as is the case for the spinoptics approach \cite{Frolov2020,Frolov_2024,PhysRevD.110.064020,Frolov_2024a,trh5-4sgq,PhysRevD.111.044001}. 

If spacetime admits a closed CKY $h_{\alpha \beta}$ and we assume $S^{\alpha \beta} p_\beta = \mathcal{O}(S^2)$ and $\dot{t}_\alpha = \mathcal{O}(S)$, then the quantity
\begin{equation}
    E = S^{\alpha \beta} h_{\alpha \beta}
\end{equation}
is conserved up to error terms of $\mathcal{O}(S^2)$. 
This can be seen by taking the derivative of $E$,
\begin{align}
    \dot{E} &= \dot{S}^{\alpha \beta} h_{\alpha \beta} + S^{\alpha \beta} \dot{x}^\sigma \nabla_\sigma h_{\alpha \beta} \nonumber\\
    &= 2 p^\alpha \dot{x}^\beta h_{\alpha \beta} + \lambda S^{\alpha \beta} p^\sigma \left( g_{\sigma \alpha} \xi_\beta - g_{\sigma \beta} \xi_\alpha \right) + \mathcal{O}(S^2) \nonumber\\
    &= \frac{2}{p \cdot t} p^\alpha S^{\beta \sigma} \dot{t}_\sigma h_{\alpha \beta} + 2\lambda S^{\alpha \beta} p_\alpha \xi_\beta + \mathcal{O}(S^2) \nonumber\\
    &= \mathcal{O}(S^2).
\end{align}
The first equality follows from the definition of $E$, and the second equality is obtained by using \cref{eq:MPD_cons} and the definition \eqref{eq:h_def} (with $\nabla_{[\mu} h_{\alpha\beta]} = 0$) of a closed CKY tensor. 
Finally, all terms in the third equality are $\mathcal{O}(S^2)$ since we assumed that $S^{\alpha \beta} p_\beta = \mathcal{O}(S^2)$ and $\dot{t}_\alpha = \mathcal{O}(S)$. 

This conserved quantity has been known in the massive case \cite{rudiger1981conserved,rudiger1983conserved,Compere:2021kjz,Ramond_2025,Compere:2023alp}, but not for the massless case of the spin Hall equations \eqref{eq:SHE} with parallel transported $t_\alpha$.

\subsection{Null linear momentum}

We shall now present additional conservation laws for massless spinning particles described by \cref{eq:MPD_cons} together with the assumption that the linear momentum is approximately null, so that $H = \tfrac{1}{2} p_\alpha p^\alpha = \mathcal{O}(S^2)$. 
Of course, the system of equations \eqref{eq:MPD_cons} together with this additional assumption continues to admit the same conserved quantities $H$, $C$ and $D$ as in \cref{eq:cons_general}. 
But if we impose $H = \mathcal{O}(S^2)$ we can relax the requirement for exact Killing vectors and tensors to their conformal counterparts.
These conserved quantities are generalizations of $C$ and $D$ in \cref{eq:cons_general} associated with conformal Killing vectors and tensors, as well as CKY tensors.

\subsubsection{Spin Hall equations}
\label{sec:cons_SHE}

In addition to the restriction $H = \mathcal{O}(S^2)$ to approximately null momenta, we now introduce the additional constraint that $S^{\alpha \beta} p_\beta = \mathcal{O}(S^2)$. 
This is the case of the spin Hall equations \eqref{eq:SHE}. 
Then, the following quantities are conserved up to error terms of $\mathcal{O}(S^2)$:
\begin{subequations} \label{eq:D_cons_SHE}
\begin{align}
    \mathcal{C} &= \kappa^\mu p_\mu + \frac{1}{2} S^{\mu \nu} \nabla_\mu \kappa_\nu, \\
    \mathcal{D} &= \mathcal{K}^{\mu \nu} p_\mu p_\nu + \mathcal{L}_{\alpha \beta \mu} S^{\alpha \beta} p^\mu, \label{eq:D_CKY}
\end{align}
\end{subequations}
where $\kappa^\mu$ is a conformal Killing vector, $\mathcal{K}^{\mu \nu} = h^{\mu \rho} h\indices{_\rho^\nu}$ is a conformal Killing tensor, $h_{\alpha \beta}$ is a CKY tensor, and
\begin{align}\label{eq: L CKY}
    \mathcal{L}_{\alpha \beta \mu} &= L_{\alpha \beta \mu} -\frac{2}{3} \left( dh_{\alpha\beta\rho}h\indices{_\mu^\rho} + h\indices{_{[\alpha}^{\rho}} dh_{\beta]\mu\rho} \right), 
\end{align}
where $L_{\alpha\beta\mu}$ is defined in \cref{eq:L_def0}. However, in this case, the terms in $L_{\alpha\beta\mu}$ proportional to the metric will not contribute due to the constraint $S^{\alpha \beta} p_\beta = \mathcal{O}(S^2)$.

The conservation of $\mathcal{C}$ has been obtained before in Ref. \cite{PhysRevD.111.024034}, but we repeat the derivation for completeness. Following similar steps as in \cref{eq:dot_C}, we have
\begin{align} 
    \dot{\mathcal{C}} &= p_\mu \dot{x}^\nu \nabla_\nu \kappa^\mu + \kappa^\mu \dot{p}_\mu + \frac{1}{2} \dot{S}^{\mu \nu} \nabla_\mu \kappa_\nu + \frac{1}{2} S^{\mu \nu} \dot{x}^\alpha \nabla_\alpha \nabla_\mu \kappa_\nu \nonumber\\
    &= \lambda p^\mu p^\nu \nabla_\nu \kappa_\mu + \frac{1}{p \cdot t} p^\mu S^{\nu \alpha} \dot{t}_\alpha \nabla_\nu \kappa_\mu - \frac{1}{2} R_{\mu \nu \alpha \beta} \kappa^\mu \dot{x}^\nu S^{\alpha \beta} \nonumber\\
    &\quad + \frac{1}{p \cdot t} p^{[\mu} S^{\nu] \alpha} \dot{t}_\alpha \nabla_\mu \kappa_\nu + \frac{1}{2} R_{\rho \alpha \mu \nu} \kappa^\rho \dot{x}^\alpha S^{\mu \nu} \nonumber\\
    &\quad + \frac{1}{2} S^{\mu \nu} \dot{x}^\alpha \left( g_{\mu \nu} \nabla_\alpha + g_{\nu \alpha} \nabla_\mu - g_{\alpha \mu} \nabla_\nu \right) \sigma \nonumber \\
    &= \lambda p^\mu p^\nu \nabla_\nu \kappa_\mu + \frac{\sigma}{p \cdot t} p_\mu S^{\mu \alpha} \dot{t}_\alpha + \lambda S^{\mu \nu} p_\mu \nabla_\nu \sigma + \mathcal{O}(S^2) \nonumber \\ 
    &= \mathcal{O}(S^2).
\end{align}
The first equality follows from the definition of $\mathcal{C}$, and the second equality is obtained by using \cref{eq:MPD_cons} and the integrability condition \eqref{CKV: int con} for the conformal Killing vector. 
The third equality follows from the defining property \eqref{eq:def_CKV1} of the conformal Killing vector, together with straightforward simplifications. Finally, the last equality is obtained using $H = \mathcal{O}(S^2)$ and $S^{\alpha \beta} p_\beta = \mathcal{O}(S^2)$.

Conservation of $\mathcal{D}$ leads to exactly the same set of equations as \eqref{eq: D_cons}, with the adjustment of the Killing tensor to the conformal Killing tensor, $K^{\alpha\beta}\to \K^{\alpha\beta}$, and the modification of $L_{\alpha\beta\mu}\to \cL_{\alpha\beta\mu}$ by replacing \cref{eq:L_def0} with \cref{eq: L CKY}. 
Equation \eqref{eq: P cubed} with these modifications is clearly satisfied
 \begin{equation}
      p^{\alpha} p^\beta p^\gamma \nabla_\gamma \K_{\alpha \beta}=p^{\alpha} p^\beta p^\gamma g_{(\alpha\beta}\tilde{\K}_{\gamma)}={\cal O}(S^2).
 \end{equation}
This leads to the general observation that, as long as the remaining equations lead to terms proportional to the contractions of $p_\alpha p^\alpha$ or $S^{\alpha\beta}p_\beta$, the equations will be satisfied.

Indeed, a lengthy calculation  using the integrability conditions in \cref{sec: int cons} (and facilitated with the {\sf xAct} package \cite{xAct} for {\sf Wolfram Mathematica} \cite{Mathematica}) leads to the following expressions \cite{suppmat}:
 \begin{subequations}
\begin{align}
      &\nabla_{(\beta} \cL_{|\mu \nu| \alpha)} - R_{\sigma (\alpha |\mu \nu|} \K\indices{^\sigma_{\beta)}}= I_{(\alpha\beta) [\mu}{}^\rho h_{|\rho| \nu]} \nonumber \\
      &\qquad\qquad-X_{\mu\nu}g_{\alpha\beta} +\frac{1}{2}(Y_{\mu(\alpha}g_{\beta)\nu}-Y_{\nu(\alpha}g_{\beta)\mu}) ,\\
      &\nabla_\mu \K_{\alpha \beta} - 2 \cL_{\mu (\alpha \beta)}=2\xi^\rho h_{\rho \mu}g_{\alpha\beta},\label{eq: K L CKY}
\end{align}
 \end{subequations}
where $I_{\alpha\beta \gamma \delta}=0$ from \cref{eq: Curv Condtions} and
\begin{subequations}
\begin{align}
   X_{\mu\nu}&=d\xi_{[\mu}{}^\rho h_{|\rho|\nu]}+\frac{2}{3}(dh_{\mu\nu\rho}\xi^\rho-3h_{[\mu}{}^\rho P_{|\rho|\nu]}),
   \\
   Y_{\mu\nu}&=d\xi_\mu{}^\rho h_{\rho\nu}-d\xi_{\nu}{}^\rho h_{\rho\mu}.
\end{align}
\end{subequations}
Here, $P_{\alpha\beta}$ is as in \cref{eq: h Ricc rel,eq: KV obstr}. 
Moreover, when $h_{\mu\nu}$ is closed, $X_{\mu\nu}$ and $Y_{\mu\nu}$ vanish identically~\cite{Frolov2017}.
It is then clear that the remaining terms in $\dot {\cal D}$ vanish when imposing $H={\cal O}(S^2)$ and $S^{\mu\nu}p_\mu={\cal O}(S^2)$,
\begin{subequations}
    \begin{align}
     S^{\mu \nu} p^\alpha p^\beta (\nabla_\beta \cL_{\mu \nu \alpha} - R_{\sigma \alpha \mu \nu} \K\indices{^\sigma_\beta})&={\cal O}(S^2),
      \\
     S^{\mu \nu} p^\alpha p^\beta \dot{t}_\nu ( \nabla_\mu \K_{\alpha \beta} - 2 \cL_{\mu \alpha \beta})&={\cal O}(S^2).
    \end{align}
\end{subequations}

Finally, it is worth emphasizing again that this calculation holds for any $\cal D$ constructed, as indicated, from a CKY $h$ and its derivative $\xi$, however no further assumption was imposed on $\xi$. 
Therefore, such a $\cal D$ will be conserved in any type D spacetime.
Moreover, one could equally have constructed $\cal D$ from the dual objects $f=\star h$ and $\zeta$ although this will not typically be an independent constant of motion.

\section{Integrability} \label{sec:integrability}

In this section, we discuss the integrability of the spin Hall equations \eqref{eq:SHE} in four-dimensional spacetimes that admit two Killing vectors $k^\mu$ and $\ell^\mu$ and a CKY tensor $h_{\mu \nu}$. Furthermore, we assume that the Killing vectors are orthogonal and linearly independent and that the Lie commutation relations in \cref{eq: KVT SN comm,eq: h Lie deriv} hold. This covers the Pleba\'nski--Demia\'nski class of spacetimes with aligned matter and the Ovcharenko--Podolsk\'y spacetimes with nonaligned matter. We expect this to also hold more generally for all type D spacetimes. However, to show the functional independence of the constants of motion, we need the specific line element \eqref{eq: PD met} with arbitrary metric functions $\Omega(r,y)$, $\Delta_r(r)$ and $\Delta_y(y)$.

For the case of massive spinning particles, the MPD equations under the Tulczyjew--Dixon SSC (and also including terms quadratic in spin) have been shown to be completely integrable in Kerr spacetimes \cite{Ramond_2025}.  
This result follows the definition of complete integrability in the sense of Liouville and Arnold \cite{JMPA_1855_1_20__137_0,arnold2006} (see also Refs. \cite{Frolov2017,PhysRevD.78.064028} for the definition of complete integrability in a general relativistic context). Furthermore, a perturbative notion of integrability \cite{PhysRevD.103.064066,fasano2006analytical} has been used in \cite{Ramond_2025}. This is unavoidable in all discussions of spinning particles in general relativity, since the MPD equations are viewed as providing small corrections to geodesic motion and terms of a certain high order in the multipole expansion need to be ignored to close the system. Also, as discussed in the previous section, some conservation laws only hold approximately.

Focusing now on the case of massless spinning particles described by the spin Hall equations \eqref{eq:SHE}, there are two important aspects regarding the definition of integrability that will be adopted here. 
First, since the spin Hall equations are defined in an approximate sense, with terms quadratic in spin and higher order multipole moments ignored, complete integrability will also be adopted in a perturbative sense, as discussed in \cite{PhysRevD.103.064066,fasano2006analytical}.
The second aspect, which is specific to the massless case and different from Ref. \cite{Ramond_2025}, is that some of the quantities introduced in \cref{sec:cons_SHE} are only conserved on shell where $H = \mathcal{O}(S^2)$. 

This does not fit with the usual definition of complete integrability \cite{Frolov2017,PhysRevD.78.064028} which requires conserved quantities in some open region of phase space. 
Note that this is also a problem when discussing the integrability of null geodesics in type D spacetimes that only possess a conformal Killing tensor.
To overcome this difficulty, we adopt here the weak form of complete integrability introduced in Ref. \cite{SARLET198587} (see also Ref. \cite{Escobar-Ruiz_2024}). 

We consider a Hamiltonian system defined on a $2d$-dimensional symplectic manifold $\mathcal{N}$ by a smooth Hamiltonian function $H$ and a symplectic $2$-form $\Omega$. 
The system is said to satisfy weak complete integrability if \cite{SARLET198587}, 
\begin{enumerate}[(i)]
    \item There exist $d$ first integrals of motion $F_i$ for which the Poisson bracket with the Hamiltonian $H = F_1$ vanishes on the subspace $H = 0$ \label{integrability_1}
    \begin{equation}
        \{F_i, H\} \big|_{H=0} = \Omega(d F_i, d H) \big|_{H=0} = 0.
    \end{equation}
    \item The integrals of motion $F_i$ are functionally independent. 
    This means that, given level sets $\mathcal{L}_\Phi$ which are subsets of $\mathcal{N}$ defined by fixing the conserved quantities $(F_1, ... , F_d) = (\Phi_1, ..., \Phi_d)$, where $\Phi_i$ are constants and $\Phi_1 = 0$, the one-forms $(d F_1, ..., d F_d)$ are linearly independent at each point of $\mathcal{L}_\Phi$. 
    In this case $\mathcal{L}_\Phi$ is a $d$-dimensional submanifold of $\mathcal{N}$. \label{integrability_2}
    \item The first integrals of motion $F_i$ are in weak involution: \label{integrability_3}
    \begin{equation}
        \{ F_i, F_j \} \big|_{H=0} = 0, \qquad \forall \, i, j.
    \end{equation}
\end{enumerate}
If the above conditions for weak complete integrability are satisfied, then the integration of the system along the hypersurface $H = 0$ is reduced to solving a single first-order ordinary differential equation \cite{SARLET198587}. We adopt this definition of weak complete integrability in a perturbative sense \cite{PhysRevD.103.064066,fasano2006analytical} and only require that the above conditions are satisfied up to error terms of $\mathcal{O}(S^2)$.

The spin Hall equations \eqref{eq:SHE} can be formulated as a Hamiltonian system over the eight-dimensional symplectic manifold $T^*M$ \cite{GSHE2020,GSHE_Dirac}, where $M$ is spacetime. 
In contrast to the massive MPD case, the spin tensor is not a dynamical variable, and we do not need a larger phase space as in Ref. \cite{Ramond_2025}. 
If we denote coordinates on $T^*M$ by $z^I = (x^\mu, p_\mu)$, we can write the Hamiltonian
\begin{equation} \label{eq:Hamiltonian}
    H(x, p) = \frac{1}{2} g^{\mu \nu}(x) p_\mu p_\nu.
\end{equation}
Then, the spin Hall equations \eqref{eq:SHE}, up to error terms of $\mathcal{O}(S^2)$, are given by
\begin{subequations}
\begin{align}
    \frac{d x^\mu}{d \tau} &= \{ x^\mu, H \} \big|_{H=\mathcal{O}(S^2)} + \mathcal{O}(S^2), \\
    \frac{d p_\mu}{d \tau} &= \{ p_\mu, H \} \big|_{H=\mathcal{O}(S^2)} + \mathcal{O}(S^2),
\end{align}
\end{subequations}
where the Poisson bracket is defined as
\begin{subequations} \label{eq:Poisson}
\begin{align}
    &\{F, G\} \big|_{H=\mathcal{O}(S^2)} = B^{I J} \frac{\partial F}{\partial z^I} \frac{\partial G}{\partial z^J} \bigg|_{H=\mathcal{O}(S^2)}, \\
    &B = \begin{pmatrix}   -\left( B_{p p} \right)^{\nu \mu} & \delta^\mu_\nu +  \left(B_{p x} \right)\indices{_\nu^\mu} \\
    -\delta^\nu_\mu +  \left(B_{x p} \right)\indices{^\nu_\mu} & -  \left(B_{x x} \right)_{\nu \mu} \end{pmatrix}.
\end{align}
\end{subequations}
The components of this tensor can be related to the components of a Berry curvature tensor and are defined as \cite{GSHE2020,HarteOancea}
\begin{subequations} \label{eq:Berry_curvature}
\begin{align}
    &\left({B_{p p}}\right)^{\beta \alpha} = \frac{ S^{\alpha \beta} }{(p \cdot t)^2} , 
    \\
    &\left({B_{xx}}\right)_{\beta \alpha} = \frac{S^{\gamma\lambda}}{2}  \bigg\{ R_{\gamma  \lambda \alpha \beta}   \nonumber \\ 
    & \qquad  +\frac{2}{ (p \cdot t)^2}  p_\rho \Gamma^\rho_{\gamma [ \alpha } \Big[ \Gamma^\sigma_{\beta] \lambda} p_\sigma - 2 (p \cdot t)  \nabla_{\beta]} t_\lambda  \Big] \bigg\},
    \\
    &\left({B_{x p}}\right)\indices{^\alpha_\beta} = - \left({B_{p x}}\right)\indices{_\beta^\alpha} = \frac{ S^{\alpha\gamma}  }{(p \cdot t)^2}  \left[ p_\rho \Gamma^\rho_{\beta \gamma} -( p \cdot t ) \nabla_\beta t_\gamma \right].
\end{align}
\end{subequations}
The Poisson bracket can also be expressed in a more compact way by using vertical and horizontal covariant derivatives (see \cite{sharafutdinov2012integral,andersson2025} for more details about the definition and properties),
\begin{subequations}
\begin{align}
    \hnabla_\mu F &= \nabla_\mu F + \Gamma^\sigma_{\mu \nu} p_\sigma \frac{\partial}{\partial p_\nu} F, \\
    \vnabla^\mu F &= \frac{\partial}{\partial p_\mu} F.
\end{align}
\end{subequations}
Using these definitions, we obtain
\begin{align}
    &\{F, G\} \big|_{H=\mathcal{O}(S^2)} = \hnabla_\alpha F \, \vnabla^\alpha G - \vnabla^\alpha F \, \hnabla_\alpha G \nonumber \\
    &- \frac{1}{(p \cdot t)^2} S^{\alpha \beta} \hnabla_\alpha F \, \hnabla_\beta G - \frac{1}{2} R_{\alpha \beta \mu \nu} S^{\mu \nu} \vnabla^\alpha F \, \vnabla^\beta G \nonumber\\
    &+ \frac{S^{\alpha \mu} \nabla_\beta t_\mu}{p \cdot t}  \left( \hnabla_\alpha F \, \vnabla^\beta G - \vnabla^\beta F \, \hnabla_\alpha G \right) \bigg|_{H=\mathcal{O}(S^2)},
\end{align}
where it is understood that the derivatives of $F,G$ are taken before the constraint is applied.

To make full use of these horizontal and vertical derivatives that act on the spin tensor, we can introduce the following tetrad~\cite{HarteOancea} $\{ t^\alpha, p^\alpha, m^\alpha, \bar{m}^\alpha \}$, where
\begin{equation}\label{eq: tet rels}
    p^\alpha t_\alpha=-\varepsilon\omega,\;  p_\alpha p^\alpha=p^2, \; t^\alpha t_\alpha=-1,\; m^\alpha\bar m_\alpha=1,
\end{equation}
and all other inner products are zero. 
Since we are working on the full phase space $T^*M$, $p_\mu$ is arbitrary and not necessarily null. 
The completeness relation for the metric is then
\begin{align}\label{eq: met tet}
    g_{\alpha\beta} &=\frac{1}{p^2+(\varepsilon\omega)^2} \big( p_{\alpha}p_{\beta}-2 \varepsilon\omega t_{(\alpha}p_{\beta)}-p^2 t_{\alpha}t_\beta \big) \nonumber\\
    &\qquad+2m_{(\alpha}\bar m_{\beta)}.
\end{align}
Of course, later, we impose the equations of motion and the Hamiltonian constraint
\begin{equation}
    H=\frac{1}{2}g^{\mu\nu}p_\mu p_\nu = \mathcal{O}(S^2) \quad \iff\quad p^2 = \mathcal{O}(S^2).
\end{equation}

As the off-shell spin tensor is orthogonal to $p_\alpha$ and $t_\alpha$, the two vectors $m^\alpha$ and $\bar m^\alpha$ describe its rotation plane so that it takes the form~\cite{HarteOancea} 
\begin{equation}\label{eq: Sm}
    \mathbb{S}^{\alpha\beta}= 2i\varepsilon s \bar{m}^{[\alpha} m^{\beta]}=\frac{\varepsilon s \epsilon^{\alpha\beta\gamma\delta}}{\sqrt{p^2+(p\cdot t)^2}}
    p_\gamma t_\delta.
\end{equation}
Notice that this is equivalent to \cref{eq:S_SHE} when $p_\mu$ is on shell. In order to correctly calculate the horizontal and vertical derivatives, one needs to keep things off shell until after differentiating.

The advantage of the tetrad is that we can project the vertical and horizontal derivatives of the tetrad onto the tetrad itself and thereby obtain the horizontal/vertical derivatives of the spin tensor (see \cite[App. A.3]{Oanceathesis} for details).
By construction, we have
 \begin{equation}
     \hnabla_\alpha p_\beta=0, \quad \vnabla^\alpha p_\beta=\delta^\alpha_\beta,
 \end{equation}
and the vertical and horizontal derivatives commute and are metric compatible
\begin{equation}
     [\vnabla,\hnabla]=0,\quad \vnabla^\mu g_{\alpha\beta}=0=\hnabla^\mu g_{\alpha\beta},
 \end{equation}
Moreover, since the timelike vector field $t^\beta$ is independent of the vertical part of phase space, we obtain
 \begin{equation}
     \vnabla^\alpha t^\beta=0.
 \end{equation}
We find, on shell of the Hamiltonian constraint, that
\begin{subequations} \label{eq:der_S}
\begin{align}
    \hnabla_\mu \mathbb{S}^{\alpha\beta} \big|_{H = \mathcal{O}(S^2)} &= \frac{2}{(p\cdot t)} p^{[\alpha}S^{\beta]\gamma}\nabla_{\mu}t_\gamma + \mathcal{O}(S^2),  \label{eq: hder S}\\
    \vnabla^\mu \mathbb{S}^{\alpha\beta} \big|_{H= \mathcal{O}(S^2)}
     &= \frac{2}{(p\cdot t)} S^{\mu[\alpha}\left(t^{\beta]}+\frac{p^{\beta]}}{p\cdot t}\right) + \mathcal{O}(S^2). \label{eq: vder S}
\end{align}    
\end{subequations}
Combining this result with \cref{eq: xdot rel}, it follows immediately that we recover the equation of motion \eqref{MPS} for the spin tensor
\begin{align}
    \dot{S}^{\mu\nu} &=\left\{ \mathbb{S}^{\mu\nu},H\right\} \big|_{H= \mathcal{O}(S^2)} \nonumber\\
    &=\frac{2}{p\cdot t}p^{[\mu} S^{\nu]\alpha}p^{\beta}\nabla_\beta t_\alpha + \mathcal{O}(S^2)
    \nonumber\\
    &=2p^{[\mu}\dot x^{\nu]} +\mathcal{O}(S^2).
\end{align}

We now check that the required conditions for weak complete integrability, as defined above, are satisfied. 
Since the phase space is eight dimensional, we only need four integrals of motion. 
As shown in \cref{sec:Cons}, in spacetimes with two Killing vectors $k^\mu$ and $\ell^\mu$ and a CKY tensor $h_{\alpha \beta}$, the integrals of motion are $\{H, C_k, C_\ell, \mathcal{D}\}$, where $H$ is the Hamiltonian in \cref{eq:Hamiltonian}, $C_\ell$, $C_\ell$ are defined as in \cref{eq: C def}, and $\mathcal{D}$ is defined as in \cref{eq:D_CKY}. 
In particular, using the Poisson bracket defined in \cref{eq:Poisson,eq:Berry_curvature}, it is straightforward to show that 
\begin{equation}
    \{F, H\} \big|_{H = \mathcal{O}(S^2)} = \dot{F}\big|_{H = \mathcal{O}(S^2)} + \mathcal{O}(S)^2 = \mathcal{O}(S^2),
\end{equation}
where $F \in \{H, C_k, C_\ell, \mathcal{D}\}$. 
Thus, the weak complete integrability condition (\ref{integrability_1}) is satisfied in a perturbative sense, up to error terms of $\mathcal{O}(S^2)$.

To satisfy the integrability condition (\ref{integrability_2}), note that it is enough to have functional independence of the integrals of motion at leading order in $S$. This can be understood in the following way. We have the set of one-forms $dF_i = (dF_0)_i + (dF_1)_i + \mathcal{O}(S^2)$, where $(dF_0)_i = \mathcal{O}(1)$ and $(dF_1)_i = \mathcal{O}(S)$. If the leading order components $(dF_0)_i$ are linearly independent, then we can build arbitrary linear combinations (up to error terms quadratic in spin)
\begin{align}
    &c^i dF_i \nonumber\\
    &= \Big[ (c_0)^i + (c_1)^i + \mathcal{O}(S^2) \Big] \Big[ (dF_0)_i + (dF_1)_i + \mathcal{O}(S^2) \Big] \nonumber\\
    &= (c_0)^i (dF_0)_i + \Big[ (c_1)^i (dF_0)_i + (c_0)^i (dF_1)_i \Big] + \mathcal{O}(S^2),
\end{align}
where $c^i$ are the coefficients of the linear combination with $(c_0)^i = \mathcal{O}(1)$ and $(c_1)^i = \mathcal{O}(S)$. Thus, the term of $\mathcal{O}(S)$ in the above equation can be controlled by $(c_1)^i (dF_0)_i$ regardless of whether $(dF_1)_i$ are linearly independent or not.

We now proceed to show that the integrals of motion are linearly independent at leading order in $S$. In other words, this is just the proof of linear independence that one would perform in the null geodesic case. The calculation is similar in spirit to the case of timelike geodesics in the Carter spacetime discussed in Ref. \cite{Krtous:2007xf}.
To show this, we consider the Pleba\'nski--Demia\'nski metric defined in \cref{sec:PD_spacetime}, working off shell with arbitrary metric functions $\Omega(r,y)$, $\Delta_r(r)$ and $\Delta_y(y)$ (this covers the Ovcharenko--Podolsk\'y spacetimes as well~\cite{Gray:2025lwy}). Furthermore, we also restrict our attention to regions of spacetime where these metric functions are nonzero and finite.  
In this case, the two Killing vectors are $k = \partial_\tau$ and $\ell = \partial_\psi$, and the CKY tensor is defined in \cref{eq:CKY_PD}. 
In canonical coordinates $(x^\mu, p_\mu) = (\tau, r, y, \psi, p_\tau, p_r, p_y, p_\psi)$, the phase space one-forms $dF_i$ are
\begin{subequations}
\begin{align}
    &dH = \frac{1}{2} p_\alpha p_\beta \left( \partial_\mu g^{\alpha \beta} \right) dx^\mu + p_\alpha g^{\alpha \mu} dp_\mu, \\
    &dC_k = k^\mu dp_\mu = dp_\tau, \\
    &dC_\ell = \ell^\mu dp_\mu = dp_\psi, \\
    &d\mathcal{D} = p_\alpha p_\beta \left( \partial_\mu \mathcal{K}^{\alpha \beta} \right) dx^\mu + 2 p_\alpha \mathcal{K}^{\alpha \mu} dp_\mu.
\end{align}
\end{subequations}
These are linearly independent on the $H = 0$ subset of $T^*M$ if
\begin{equation} \label{eq:lin_indep}
    dC_k \wedge dC_\ell \wedge dH \wedge d\mathcal{D} \big|_{H=0} \neq 0.
\end{equation}
To simplify the calculation, we introduce the notation
\begin{subequations} \label{eq:A_B}
\begin{align}
    \mathcal{A}(r, p_r) &= \Delta_r p_r^2 - \frac{(C_\ell + r^2 C_k)^2}{\Delta_r}, \\
    \mathcal{B}(y, p_y) &= \Delta_y p_y^2 + \frac{(C_\ell - y^2 C_k)^2}{\Delta_y},
\end{align}
\end{subequations}
such that the Hamiltonian and Carter's constant can be written as
\begin{subequations}
\begin{align}
    H &= \frac{\Omega^2}{2 \Sigma} (\mathcal{A} + \mathcal{B}), \\
    \mathcal{D} &= \frac{1}{\Sigma} (r^2 \mathcal{A} - y^2 \mathcal{B}).
\end{align}
\end{subequations}
Note that $H = 0$ is equivalent to imposing $\mathcal{A} + \mathcal{B} = 0$. With these definitions, we can express the Hamiltonian and the Carter one-forms, as well as their wedge product as
\begin{subequations}
\begin{align}
    dH \big|_{H=0} &= \frac{\Omega^2}{2 \Sigma} (d\mathcal{A} + d\mathcal{B}) \big|_{H=0}, \\
    d\mathcal{D} \big|_{H=0} &= \frac{1}{\Sigma} (r^2 d\mathcal{A} - y^2 d\mathcal{B}) \big|_{H=0}, \\
    dH \wedge d\mathcal{D} \big|_{H=0} &= - \frac{\Omega^2}{2 \Sigma} d\mathcal{A} \wedge d\mathcal{B} \big|_{H=0}.
\end{align}
\end{subequations}
Inserting back into \cref{eq:lin_indep}, the linear independence condition becomes
\begin{equation}
    - \frac{\Omega^2}{2 \Sigma} dC_k \wedge dC_\ell \wedge d\mathcal{A} \wedge d\mathcal{B} \big|_{H=0} \neq 0.
\end{equation}
The one-forms $dC_k = dp_\tau$ and $dC_\ell = dp_\psi$ are linearly independent because the Killing vectors $k$ and $\ell$ are linearly independent. Using \cref{eq:A_B}, we have $d\mathcal{A} \in \mathrm{span}\{dr, dp_r\}$ and $d\mathcal{B} \in \mathrm{span}\{dy, dp_y\}$. Thus, the linear independence condition is satisfied unless we have
\begin{equation}
    d\mathcal{A} \big|_{H=0} = 0 \qquad \textrm{or} \qquad d\mathcal{B} \big|_{H=0} = 0.
\end{equation}
Recalling that we restricted our attention to spacetime regions where the metric functions $\Omega$, $\Delta_r$, and $\Delta_y$ are nonzero and finite, the above conditions are then equivalent to
\begin{subequations}
\begin{align}
    d\mathcal{A} \big|_{H=0} &= 0 \quad \Rightarrow \quad p_r = 0 = \partial_r \frac{(C_\ell + r^2 C_k)^2}{\Delta_r}  \\
    d\mathcal{B} \big|_{H=0} &= 0 \quad \Rightarrow \quad p_y = 0 = \partial_y \frac{(C_\ell - y^2 C_k)^2}{\Delta_y}.
\end{align}
\end{subequations}
These conditions correspond to constant-$r$ and constant-$y$ null geodesic motion, as can easily be seen by examining the Hamiltonian form of the geodesic equations,
\begin{subequations}
\begin{align}
    \dot{r} &= \frac{\partial H}{\partial p_r} \bigg|_{H=0} = \frac{\Omega^2}{\Sigma} \Delta_r p_r = 0, \\
    \dot{p}_r &= -\frac{\partial H}{\partial r} \bigg|_{H=0} = \frac{\Omega^2}{2 \Sigma} \frac{\partial \mathcal{A}}{\partial r} = 0,
\end{align}
\end{subequations}
and similarly for $\dot{y}$ and $\dot{p}_y$. Thus, we conclude that the integrals of motion are functionally independent and the integrability condition (\ref{integrability_2}) is generally satisfied, provided that the metric functions are nonzero and finite and that we exclude the special cases of constant-$r$ and constant-$y$ null geodesic motion.

Finally, we now check that the constants of motion are in weak involution for the integrability condition (\ref{integrability_3}) to be satisfied. Using \cref{eq:der_S}, the Poisson bracket between two constants of motion arising from two Killing vectors $k$ and $\ell$ reads
\begin{align}\label{eq: C com}
    &\left\{ C_k, C_\ell \right\} \big|_{H=\mathcal{O}(S^2)} = -p\cdot[k, \ell] + \frac{1}{2}S^{\alpha\beta} k^\mu \ell^\nu R_{\alpha\beta\mu\nu} \nonumber\\
     &-\frac{S^{\alpha\beta}}{(p\cdot t)^2} \left[ {p^\mu p^\nu}+2\,(p\cdot t) p^{(\mu} t^{\nu)}\right] \nabla_\alpha k_\mu \nabla_\nu \ell_\beta +\mathcal{O}(S^2) \nonumber\\
     &=-p\cdot[k, \ell] + \frac{1}{2}S^{\alpha\beta} k^\mu \ell^\nu R_{\alpha\beta\mu\nu} - S^{\alpha\beta}\nabla_{\alpha} k_\lambda \nabla^{\lambda} \ell_{\beta} 
     \nonumber\\
     &\qquad+\mathcal{O}(S^2)
     \nonumber\\
     &= \mathcal{O}(S^2),
\end{align}
where we have used the Killing vector integrability condition \eqref{CKV: int con}.
The second equality follows from the relation between our tetrad and the metric in \cref{eq: met tet}, with $p^2 = \mathcal{O}(S^2)$, and the expression for $\mathbb{S}^{\alpha\beta}$ in \cref{eq: Sm}. 
The final equality follows most easily by considering Killing coordinates where $k = \partial_\tau$ and $\ell = \partial_\psi$ \cite[Eq. (295)]{Compere:2021kjz}. 
Thus, we have the desired commutativity.

Next, we consider the Poisson bracket between $C_k$, and $\mathcal{D}$. Using the Killing vector equation and the respective integrability condition in \cref{CKV: int con}, we obtain
\begin{align}
    &\{C_k, \mathcal{D}\} \big|_{H=\mathcal{O}(S^2)} = -[k, \mathcal{K}]^{\mu\nu} p_\mu p_\nu \nonumber\\
    & -\frac{1}{p\cdot t}p_\alpha p_\beta S^{\mu\nu}(k^\rho\nabla_\rho t_\nu+ t^\rho\nabla_\rho k_\nu)\left[\nabla_\mu \mathcal{K}^{\alpha\beta}-2\mathcal{L}_\mu{}^{\alpha\beta} \right] \nonumber\\
    &-2p_\alpha S^{\beta\gamma}(k^\rho\nabla_\rho \mathcal{L}_{\beta\gamma}{}^\alpha-\mathcal{L}_{\beta\gamma}{}^\rho\nabla_\rho k^\alpha) \nonumber\\
    &+\frac{4}{p\cdot t}p^\alpha p^\beta t^\gamma S^{\delta\epsilon}(\mathcal{L}_{\gamma\delta\alpha}\nabla_\beta k_\epsilon+ \mathcal{L}_{\delta\alpha\beta}\nabla_{\gamma}k_\epsilon) \nonumber\\
    &-\frac{2}{(p\cdot t)^2}p_\alpha p_\beta  p_\gamma S^{\delta\epsilon}(\nabla_\epsilon \mathcal{K}^{\beta \gamma}+ \mathcal{L}^{\beta}{}_{\epsilon}{}^{\gamma})\nabla_\delta l^\alpha+\mathcal{O}(S^2) \nonumber\\
    &= -[k, \mathcal{K}]^{\mu\nu}p_\mu p_\nu - 2 S^{\alpha\beta}p^{\gamma} [k, \mathcal{L}]_{\alpha\beta\gamma} + \mathcal{O}(S^2) \nonumber\\
    &= \mathcal{O}(S^2).
\end{align}
Here, $[k, \cdot]$ denotes the Lie derivative of a tensor field along the vector $k$. The second equality follows from our previously derived \cref{eq: K L CKY} for $\mathcal{L}_{\mu\alpha\beta}$, and using the on-shell form of the metric in \cref{eq: met tet}.
The final equality uses the Lie commutation of the Killing vector and the conformal Killing tensor, \cref{eq: KVT SN comm}, and $h$, \cref{eq: h Lie deriv}, for the class of spacetimes of interest.
Notice that each step in this calculation applies equally to $\ell$, hence we conclude 
\begin{equation}
    \{C_\ell,{\cal D}\}={\cal O}(S^2).
\end{equation}

Thus, we have obtained the required commutation relations between the constants of motion such that the set $\{H,C_k, C_\ell, {\cal D}\}$ is in weak involution. Therefore, we have demonstrated that, in the Pleba\'nski--Demia\'nski or the Ovcharenko--Podolsk\'y classes of type D spacetimes, the spin Hall equations are completely integrable in the weak sense of Ref. \cite{SARLET198587}.

\section{Conclusions} \label{sec:conclusions}

To summarize, we have constructed conserved quantities in a large class of type D spacetimes for massive and massless spinning particles subject to the Mathisson--Papapetrou--Dixon equations of motion and the spin Hall specialization, respectively.
In doing so, we have emphasized the role of the conformal Killing--Yano tensors inherent to type D spacetimes, and we have shown that these give rise to conservation laws that generalize Carter's constant for spinning particles.

New to this work, in the massive case, we have shown that the generalized Carter constant $D$ in \cref{eq: D def} built from the closed conformal Killing--Yano tensor is conserved independently of the spin supplementary condition imposed. Previously, this conservation law has only been obtained under the assumption of the Tulczyjew–Dixon spin supplementary condition $S^{\alpha \beta} p_\beta = 0$.
For massless spinning particles, the existence of a generalized Carter constant has not been obtained before, so the conservation of $D$ represents a new result. Furthermore, we also extended this to the case where spacetime only admits a conformal Killing--Yano tensor, leading to the conserved quantity $\cal D$ in \cref{eq:D_CKY}. Our general strategy to obtain these conserved quantities, similar to previous approaches for the massive case \cite{rudiger1981conserved,rudiger1983conserved}, was to write the unknown tensors $L_{\alpha \beta \mu}$ in \cref{eq: D def} and $\mathcal{L}_{\alpha \beta \mu}$ in \cref{eq:D_CKY} as the most general possible combination of the conformal Killing–Yano tensor $h$ and its Killing vector derivative $\xi_\mu=\frac{1}{3}\nabla^\nu h_{\nu\mu}$, and then constrain the possible terms by requiring conservation.

Using these results, we have also demonstrated the integrability of the spin Hall equations in the off-shell spacetimes conformally related to Carter spacetimes, which encompasses the Carter, Pleba\'nski--Demia\'nski and Ovcharenko--Podolsk\'y classes of black-hole spacetimes.
This required a perturbative and weak notion of integrability, which extends the usual notion to one where integration takes place on a subspace of $T^*M$ corresponding to the null constraint, $H=\frac{1}{2}g^{\mu\nu}p_\mu p_\nu={\cal O}(S^2)$, for massless particles. Furthermore, if we ignore the spin-dependent effects by setting $S^{\mu \nu} = 0$, then our result reduces to a proof of (nonperturbative) weak complete integrability for null geodesics in this large class of type D spacetimes.

There are several interesting avenues opened up by our results. 
Namely, integrability following from the constants of motion was demonstrated for a large class of type D spacetimes.
However, only the functional independence of this set required imposing a particular line element. 
It should be possible to relax this restriction so that our results apply to all type D spacetimes.
Next, these conserved quantities were built from the conformal symmetries of the spacetime, so it is natural to expect that the spin Hall equations have some underlying conformal symmetry too, at least in type D spacetimes. 
This has been demonstrated in the spinorial language for test fields \cite{Araneda:2018ezs}, so one might seek to relate these results by writing the spin Hall equations in that framework.
Then, we also required imposing the constraint $S^{\mu\nu} p_\nu={\cal O}(S^2)$, which is valid for particular types of wave packets with longitudinal angular momentum. It remains to be seen whether this can be relaxed.
Moreover, as is clear from our work, these results hold perturbatively in spin, but can one go beyond this linear approximation? Such extensions are known in the massive case \cite{Ramond_2025}, and it is generally known that there is a subtle interplay between the hidden symmetries of spacetime, represented by conformal Killing--Yano tensors, and higher-order multipole moments in the Mathisson--Papapetrou--Dixon equations \cite{PhysRevD.108.124005}.  

Finally, and perhaps most importantly, the results should facilitate the study of wave propagation phenomena and the resulting observational implications for this very general class of black holes \cite{GSHE_lensing,GSHE_lensing2,Frolov_2024a,trh5-4sgq,PhysRevD.110.124011}. We leave these directions to future study.

\section*{Acknowledgments}

The work of L.A. was funded in part by NSFC under grant W2431012. This research was funded in whole or in part by the Austrian Science Fund (FWF) \href{https://doi.org/10.55776/PIN9589124}{10.55776/PIN9589124}.

\bibliography{references}

\end{document}